\documentclass[11pt]{article}
\begin{document}
\title
{On the Landau system in noncommutative phase-space}
\author{
{\bf {\normalsize Sunandan Gangopadhyay}$^{a,c}$\thanks{sunandan.gangopadhyay@gmail.com}},
{\bf {\normalsize Anirban Saha}$^{a,c}$\thanks{anirban@iucaa.ernet.in}}\\[0.2cm]
$^{a}$ {\normalsize Department of Physics, West Bengal State University, Barasat, Kolkata 700126, India}\\[0.2cm]
{\bf {\normalsize Aslam Halder}$^{b}
$\thanks{aslamhalder.phy@gmail.com}},\\
$^{b}$ {\normalsize Kolorah H.A.W Institution, Kolorah, Howrah-711411, India  }\\[0.2cm]
$^{c}${\normalsize Visiting Associate in Inter University Centre for Astronomy $\&$ Astrophysics (IUCAA),}\\
{\normalsize Pune, India}\\[0.3cm]
}
\date{}

\maketitle
\begin{abstract}
\noindent
We consider a charged particle moving in a two dimensional plane in the presence of a background magnetic field perpendicular to the plane, i.e. the Landau system in a phase-space where the coordinates and momenta both follow canonical noncommutative algebra. A set of generalized transformations is derived in this paper which maps the NC problem to an equivalent commutative problem. In this set up, we study the Aharonov-Bohm effect and the Landau levels. For the Aharonov-Bohm effect, the phase-shift is found to contain corrections due to phase-space noncommutativity and also depends on the scaling parameter appearing in the generalized transformations.
The result agrees with those in the literature upto first order in the noncommutative parameters when proper choice of the scaling parameter is taken. We then obtain the magnetic length and degeneracy of the Landau levels, both are seen to admit NC corrections. The Landau levels are seen to get altered due to phase-space noncommutativity as well. This energy spectrum of the Landau system is computed from two different perspectives, namely the explicit NC variable approach and the commutative-equivalent approach. The results match exactly, solidifying the evidence in favour of the equivalence of the two approaches. 
\end{abstract}

\maketitle

\vskip 1cm
\section{Introduction}
In 1930, Landau analyzed the quantum dynamics of a charged particle moving in a background homogeneous magnetic field (to be referred to as the Landau system hereafter) to show that it poses quantized energy levels \cite{landau}. These quantized energy levels, dubbed the Landau levels (LL's), arise in a plethora of important physical scenarios- the integral and fractional quantum Hall effect \cite{ref_1, laugh_1}, Aharonov-Bohm effect \cite{aharonov}, different two-dimensional surfaces \cite{ref_2, ref_3} like Graphene \cite{Novoselov, Zhang}, anyons excitations in a rotating Bose–Einstein condensate \cite{ref_4, ref_5} etc are to name but a few. 
Apart from these wide-spread occurence, on a more formal note, the Landau problem is perhaps the prototypical example of space quantization where one arrives at a coordinate space following a noncommutative algebra, in rudimentary quantum mechanics. 

To briefly review this intriguing behaviour let us consider
 a charged particle of mass $m$ moving in the plane $\vec x=(x_1,x_2)$ in presence of a constant, perpendicular magnetic field $B$. The Lagrangian will be 
\begin{equation}
{L}=\frac {m}{2}\dot{\vec x}^2- e\dot{\vec x}\cdot\vec A
\label{Landaum}
\end{equation}
with the vector potential in the symmetric gauge given by\footnote{We follow the convension $\epsilon_{ij}=-\epsilon_{ji}$, with $\epsilon_{12}=1$.} 
\begin{eqnarray} 
A_i=-\frac B2\,\epsilon_{ij}\,x_j
\label{c_sym_Gauge}
\end{eqnarray}
The Hamiltonian can be written in terms of the the gauge invariant observable mechanical momentum $\vec\pi = m \dot{\vec x} = \vec p + e \vec A$ as
\begin{eqnarray}
{H}=\frac1{2m}\,\vec\pi^{\,2}
\label{e1}
\end{eqnarray} 
Note that $\vec p$ is the canonical momentum that may vary with gauge choice. Upon quantization by imposing the usual canonical commutation relations it follows that the operators corresponding to the physical momentum have the non-vanishing quantum commutators
$\left[\pi^{op}_{i}\,,\, \pi^{op}_{j}\right]=i\hbar eB \epsilon_{ij} \ ,$
showing that the physical momenta, in presence of a background magnetic field $\vec B$, belong to a noncommutative (NC) momentum space. Expressing them in terms of the harmonic oscillator creation and annihilation operators, the energy eigenvalues of the Hamiltonian are the LL's 
\begin{eqnarray}
E_{n} = \left( n + \frac{1}{2}\right) \hbar \omega_{c}
\label{LL}
\end{eqnarray}
with $\omega_{c} = (\frac{eB}{m})$, the cyclotron frequency \footnote{Velocity of light is taken $c = 1$ throughout the paper.}. In the limit $m \to 0$ with fixed $B$ or equivalently $B >> m$ the mass gap between the Landau level $\Delta \omega_{c}$ grows and consequently we get the projection of the whole spectrum onto the lowest LL.
In this limit (\ref{Landaum}) becomes a first order Lagrangian
${L}_0=-\frac B2\,\dot x_i\,\epsilon_{ij}\,x_j$
which is already expressed in phase-space with the spatial coordinates $x_1,x_2$ being the canonically conjugate variables so that
$\left[x^{op}_i\,,\, x^{op}_j\right]= i\frac{\hbar}{2B}\,\epsilon_{ij} \ .$
Thus we can conclude that noncommuting coordinates arise in electronic systems constrained to lie in the lowest Landau level.

Remarkably, a direct analogy to this simple example arises in string theory with D-branes in background ”magnetic fields”~\cite{acny}. D-brane worldvolume can be shown to become a noncommutative (NC) space and a low-energy effective field theory can be arrived at in the point particle limit, where the string length goes to zero. This is known as a noncommutative field theory (NCFT) \cite{szabo}-\cite{carroll} where the coordinate algebra induces a space-time uncertainty relation and the notion of a spacetime point is replaced by a Planck cell of dimension given by the Planck area $|\theta_{ij}|$. 
Similar NC spatial geometry is also known to arise in various theories of quantum gravity \cite{sw}-\cite{mof}. 

The low energy limit of this NCFT gives us the noncommutative quantum mechanics (NCQM) \cite{duval}-\cite{sgthesis} where, we speculate that some relic of the Planck scale effect may be traced \cite{carroll, ani, stern, galileo, ani_equi}. 
It would be indeed intriguing to see if such traces can be found in the Landau system itself which has such deep an analogy with the noncommutativity of space as we have discussed above. This problem was addressed by many authors in the literature 
\cite{Iengo}-\cite{Alvarez} from different perspectives. In recent years, there have also been speculations of a more elaborate NC phase-space structure \cite{momentum_NC2}-\cite{samanta}, so we carry out our entire analysis in NC phase-space for completeness and generality. Our results can be readily cast into the special case of configuration-space (spatial) noncommutativity only, by equating the momentum NC parameter to zero.


The primary aim of this paper is to study the Landau problem defined over the four-dimensional NC phase-space where operators corresponding to the canonical pairs, denoted by $\left(\hat{x}_i \, , \, \hat{p}_i\right)$ follow NC algebra:
\begin{eqnarray}
\label{e4}
\left[\hat{x}_{i},\hat{x}_{j}\right]=i\theta_{ij} = i\theta \epsilon_{ij} ~ ; \quad \left[\hat{p}_{i},\hat{p}_{j}\right]=i\bar{\theta}_{ij} = i\bar{\theta} \epsilon_{ij}~ ; \quad \left[\hat{x}_{i},\hat{p}_{j}\right] = i\tilde{\hbar}\delta_{ij}.
\end{eqnarray}
Here $\theta$ and $\bar{\theta}$ denots the spatial and momentum noncommutative parameters and $\tilde{\hbar}=\hbar(1+\frac{\theta\bar{\theta}}{4\hbar^2})$ is the effective Planck's constant. The usual approach in the literature to deal with such problems is to form an equivalent commutative description of the NC theory by employing some transformation which relate the NC phase-space variables (and the related operators) to ordinary commutative variables (operators) $x_{i}$ and $p_{i}$ satisfying the usual operator Heisenberg algebra 
\begin{eqnarray}
\left[{x}^{op}{}_{i}, {p}^{op}{}_{j}\right]=i\hbar \delta_{ij}~; \quad \left[x^{op}{}_{i} \, , \, x^{op}{}_{j}\right]=0= \left[p^{op}{}_{i},p^{op}{}_{j}\right].\label{cAlgebra}
\end{eqnarray}
In this paper we first carry out our investigation of the Landau system using NC variables explicitly. Specifically, we check whether the magnetic length $l_{B}$ of this system and the degeneracy of the Landau levels \cite{yan, Goerbig} acquire corrections from the NC phase-space structure. Surprisingly, these aspects of the Landau system, though very important in context of various observable effects in experimental condenced matter (e.g., the Hall effect) has not been emphasized much in the contemporary NC literature \cite{Iengo, Gamboa0, Gamboa1, Das, Nair1, Dulat, Giri, Alvarez}. We also compute the spectrum for the system, i.e., the phase-space NCLL. To verify the consistency of our results, we also work out this NC phase-space spectrum taking the usual approach, i.e., by quantizing the commutative-equivalent Hamiltonian obtained using a set of generalized transformations (which we shall derive in this paper) and confront it with the former. Reassuringly, these two NC spectra match exactly, establishing that the present description of the Landau system is unambiguous. Note that unlike the non-linear maps used in \cite{harms, Dulat, Dulat1}, the change of variables used in this paper to obtain the commutative equivalent Hamiltonian are exact maps. The NC phase-space algebra (\ref{e4}) also differs from the one used in \cite{harms, Dulat} where similar energy-spectra for the commutative-equivalent theory have been produced.

However, before delving into the analysis of the NC Landau problem, we first study the consequence of phase-space noncommutativity in another important phenomena concerning the Landau syatem, namely, the Aharonov-Bohm (AB) effect. The significance of the AB effect lies in the fact that it elevates our notion of the electromagnetic potential from being a convenient mathematical concept in Electrodynamics to a physical quantity in quantum mechanics. 
AB effect arises when one considers a beam of electrons split into two parts, moving in the vicinity of a solenoid placed perpendicular to the plane of the beam. The recombination of these two beams of electrons results in a phase-shift in the interference pattern which depends on the magnetic flux enclosed by the two alternative beam paths. This phase-shift is observed even though the electron-beams move through regions in space devoid of any magnetic field, and only having non-vanishing vector potential, thus establishing the physicality of the latter. 
Since the electromagnetic vector potential is fixed by the gauge choice for a given background magnetic field in a way (see equation (\ref{c_sym_Gauge})) that will be essentially altered by noncommutativity, it is imperative to check if the NC framework alters the observed phase-shift non-trivially. Further as we have chosen to work with the symmetric gauge in analogy with the commutative scenario in this paper, we employ the usual approach of mapping the NC Hamiltonian of the theory to an equivalent commutatative Hamiltonian (with NC corrections) to study the AB effect.

This article is organised as follows. In the next section we derive a mapping between the NC and commutative sets of variables. In section 3, we present the study of the Aharonov-Bohm effect in NC phase-space, specifically computing the AB phase.  Along the way, we describe the framework of obtaining the commutative-equivalent scenario for a theory defined over the NC phase-space. Section 4, contains the analysis of the Landau system in NC phase-space. We conclude in section 5.   
\section{Generalized mapping between noncommutative and commutative variables}
In this section, we derive a generalized mapping between the NC and commutative sets of variables \cite{Wang}. We relate the two sets of variables by the following equations
\begin{eqnarray}
\label{e74}
\hat{x}_{i}=a_{ij}x_{j}+b_{ij}p_{j}
\end{eqnarray}
\begin{eqnarray}
\label{e75}
\hat{p}_{i}=c_{ij}x_{j}+d_{ij}p_{j}
\end{eqnarray}
where $a$, $b$, $c$ and $d$ are $2\times 2$ transformation matrices. To determine the conditions that the transformation matrices
should satisfy, we use the NC algebra (\ref{e4}) and the commutative algebra (\ref{cAlgebra}), which yields
\begin{eqnarray}
\label{e76}
ad^T-bc^T=\frac{\tilde{\hbar}}{\hbar}
\end{eqnarray}
\begin{eqnarray}
\label{e77}
ab^T-ba^T=\frac{\theta}{\hbar}
\end{eqnarray}
\begin{eqnarray}
\label{e78}
cd^T-dc^T=\frac{\bar{\theta}}{\hbar}
\end{eqnarray}
where $\theta$ and $\bar{\theta}$ are $2\times 2$ antisymmetric matrices. To proceed further, we assume $a_{ij}=\alpha\delta_{ij}$ , $d_{ij}=\beta\delta_{ij}$, where $\alpha$ and $\beta$ are two scaling constants . 
With these assumptions, eq.(s) (\ref{e77}) and (\ref{e78}) give the solutions for the matrices $b$ and $c$ as
\begin{eqnarray}
\label{e79}
b_{ij}=-\frac{1}{2\alpha\hbar}\theta_{ij}
\end{eqnarray}
\begin{eqnarray}
\label{e80}
c_{ij}=\frac{1}{2\beta\hbar}\bar{\theta}_{ij}.
\end{eqnarray}
Substituting the expressions of $a$, $b$, $c$ and $d$ in eq.(\ref{e76}), we get the generalized expression for the effective Planck's constant 
\begin{eqnarray}
\label{e81}
\tilde{\hbar}=\alpha\beta\hbar\left(1+\frac{\theta\bar{\theta}}{4\hbar^2\alpha^2\beta^2}\right).
\end{eqnarray}
In the commutative limit ($\theta=0=\bar{\theta}$), we must have $\tilde{\hbar}=\hbar$. This implies that $\alpha\beta=1$ which in turn implies $\tilde{\hbar}=\hbar(1+\frac{\theta\bar{\theta}}{4\hbar^2})$.
Finally substituting the expressions of $a$, $b$, $c$ and $d$ into eq.(s) (\ref{e74}) and (\ref{e75}), we obtain the set of generalized transformations 
\begin{eqnarray}
\label{e86}
\hat{x}_{i}  =   \alpha\left(x_{i}-\frac{1}{2\hbar\alpha^2}\theta_{ij}p_{j}\right)
\end{eqnarray}
\begin{eqnarray}
\label{e8zz} 
\hat{p}_{i}  = \frac{1}{\alpha}\left( p_{i}+\frac{\alpha^2}{2\hbar}\bar{\theta}_{ij}x_{j}\right) .
\end{eqnarray}  
With the choice $\alpha=1=\beta$, the above set of tranformations reduce to the well known transformations \cite{momentum_NC2} 
\footnote{We drop the superscript `op' in the remainder of the paper as the meaning should be obvious from the context.}
\begin{eqnarray}
\label{e6}
\hat{x}_{i}  =   x_{i}-\frac{1}{2\hbar}\theta_{ij}p_{j}
\end{eqnarray} 
\begin{eqnarray}
\label{e6a} 
\hat{p}_{i}  =  p_{i}+\frac{1}{2\hbar}\bar{\theta}_{ij}x_{j} .
\end{eqnarray}
For $\bar{\theta}=0$, the transformations can be related to the Moyal star product between two functions
\begin{eqnarray}
\label{e70}
f(x)\star g(x)=f\left(x-\frac{\tilde{p}}{2\hbar}\right)g(x)
\end{eqnarray}
where $\tilde{p}=\theta p$.
The effect of this star product is to shift the argument of the function $f$ by a $\theta$ dependent factor and the argument is found to be identical to the change of variables (\ref{e6}).

Note that there are other forms of the NC phase-space algebra (different from the one employed in this paper) which have been mapped to the commutative Heisenberg algebra (\ref{cAlgebra}) by non-linear transformations \cite{harms, Dulat, Dulat1}.

\section{Aharonov-Bohm effect in noncommutative phase-space }
We consider a charge moving in a two dimensional NC plane with a background magnetic field. The Hamiltonian of the system in NC phase-space will be an immediate generalization of (\ref{e1}): 
\begin{eqnarray}
\label{e1a}
\hat{H}=\frac{1}{2m}\sum_{i}(\hat{p_{i}}+e\hat{A_{i}})^2  ~; \quad (i=1,2)
\end{eqnarray}
where the phase-space variables of (\ref{e1}) are replace by the corresponding NC phase-space operators, defined through the commutation relations (\ref{e4}).
For the subsequent analysis, we make the gauge-choice in analogy with the symmetric gauge of commutative gauge theory, namely
\begin{eqnarray}
\label{e2}
\hat{A_{i}}=-\frac{B}{2}\epsilon_{ij}\hat{x}_{j}
\end{eqnarray}
and the NC phase-space Hamiltonian (\ref{e1a}) becomes
\begin{eqnarray}
\label{e3}
\hat{H}=\frac{1}{2m}\sum_{i}(\hat{p_{i}}-\frac{eB}{2}\epsilon_{ij}\hat{x}_{j})^2.
\end{eqnarray}
Substituting the generalized transformations (\ref{e86}, \ref{e8zz}) in eq.(\ref{e3}), we get an equivalent commutative Hamiltonian in terms of the commutative phase-space variables (operators) which describes the original system defined over the NC phase-space :
\begin{eqnarray}
\label{e7}
\hat{H} & = &\frac{1}{2m} \left(a^{2}p_{i}{}^2+b^{2} x_{i}{}^2 + 2 a b \epsilon_{ki}x_{k} p_{i} \right )\\
a&=&\frac{1}{\alpha}\left(1-\frac{eB\theta}{4\hbar}\right) \quad,\quad b=\frac{eB\alpha}{2}\left(1-\frac{\bar{\theta}}{eB\hbar}\right). \nonumber
\end{eqnarray} 
Defining $\vec{p}=(\hat{i}p_{1}+\hat{j}p_{2})$ and $\vec{r}~'=(-\hat{i}x_{2}+\hat{j}x_{1})$, the Hamiltonian (\ref{e7}) can be recast in the following convenient form  
\begin{eqnarray}
\label{e8}
\hat{H}=\frac{1}{2m}(a\vec{p}-b\vec{r}~')^2.
\end{eqnarray}
To compute the AB phase-shift, we now consider the Schr\"{o}dinger equation 
\begin{eqnarray}
\label{e9}
\frac{1}{2m}\left[\left(a\vec{p}-b\vec{r}~' \right)^2\right]\psi=i\hbar\frac{\partial\psi}{\partial t}~.
\end{eqnarray}
Writing the wavefunction in the form
\begin{eqnarray}
\label{e10} 
\psi=e^{ig}\psi'
\end{eqnarray}
it is trivial to check from the above equation that the physics remains unaltered under the gauge transformation $\vec{A}\rightarrow\vec{A}+\vec{\nabla}g$. This yeilds the phase-shift to be 
\begin{eqnarray}
\label{e11}
g(\vec{r})=\oint^{r}\vec{A}'.d\vec{r} ~;~ \vec{r}=\hat{i}x_{1}+\hat{j}x_{2}
\end{eqnarray}
where $\vec{A}'$ can be identified from eq.(\ref{e9}) to be
\begin{eqnarray}
\label{e12a}
\vec{A}'=\frac{b}{\hbar a}\vec{r}~'.
\end{eqnarray}
Computing this expression, we get the AB phase
\begin{eqnarray}
\label{e13a}
g(\vec{r})&=&\oint^{r}\vec{r}~'.d\vec{r} = \frac{2\pi r^{2}}{\hbar} \left(\frac{b}{a}\right) = \frac{eB\alpha}{\hbar}\frac{(1-\frac{\bar{\theta}}{eB\hbar})}{\frac{1}{\alpha}(1-\frac{eB\theta}{4\hbar})}\pi r^{2}\nonumber\\
&=&\frac{eB\alpha^2}{\hbar}\left(1+\frac{eB\theta}{4\hbar}-\frac{\bar{\theta}}{eB\hbar}+\frac{e^{2}B^{2}\theta^{2}}{16\hbar^{2}}-\frac{\theta\bar{\theta}}{4\hbar^{2}}\right)\pi r^{2}+\mathcal{O}(\theta^{3}, \bar{\theta}^{3}, \theta^{2}\bar{\theta}, \theta\bar{\theta}^{2}).
\end{eqnarray}
The AB phase is thus seen to pick up NC correction factor that has leading order correction from both spatial and momentum noncommutativity. The result is also found to be dependent on the scale factor $\alpha$ appearing in the generalized transformations (\ref{e86}) and (\ref{e8zz}). This is also a new result in this paper. Note that owing to the extreme smallness of the NC parameters, one can only hope that the leading order corrections will have some observable effects in near-future experiments. With the present observational accuracy of AB effect, the bounds set on the NC parameters are far weaker than those set by observation of e.g., experimental test of Lorentz violation \cite{carroll}. However, the important aspect of the present result is the very different nature of the two leading order corrections coming from two different sectors of the NC algebra.
Note that while the correction from the spatial noncommutativity will grow with increasing background magnetic field, that from the momentum noncommutativity will diminish and vice-versa. This reverse character of the two leading order corrections can be used while analyzing the experimental data to identify which sector of noncommutativity (spatial or momentum) they belong to. 

With the choice $\alpha=1$, the expression (\ref{e13a}) agrees with that obtained in \cite{harms} (where non-linear transformations have been used to map a NC phase-space algebra (which differs from the algebra in this paper) to the commutative algebra) 
upto first order in $\theta$ and  $\bar{\theta}$  and reduces to the commutative result in $\theta=\bar{\theta}=0$ limit. 
In the $\bar{\theta}=0$ limit, the result to leading order in $\theta$ also agrees with that obtained by the path integral approach 
\cite{sg42} . The $\theta=0$ limit of eq. (\ref{e13a}), however, cannot be matched with the path integral formalism as it has not been developed with momentum noncommutativity.

This concludes our analysis of the AB phase. Note that the NC Hamiltonian (\ref{e3}) and its commutative-equivalent form (\ref{e7}) obtained here will be used in the next section where we will analyze the Landau problem in NC phase-space.
 
\section{Landau problem in noncommutative phase-space}
In this section our strategy is to work with NC phase-space variables first. It will be seen that this gives an elegant way of accessing the magnetic length, the  Landau-level degeneracy and also the energy spectrum. We shall further compute the spectrum using the commutative equivalent picture as well to show that they produce the same result.

We first compute the degeneracy of the Landau problem. To do this, we observe that the components of the NC mechanical momenta operators $\hat{\pi}_{x}=\hat{p}_{x}+e\hat{A}_{x}$ and $\hat{\pi}_{y}=\hat{p}_{y}+e\hat{A}_{y}$, obtained from the NC phase-space Hamiltonian (\ref{e1a}) satisfy the commutation relation
\begin{eqnarray}
\label{e15}
\left[\hat{\pi}_{x},\hat{\pi}_{y}\right ]=-i\left(eB\hbar-\frac{e^2B^2}{4}\theta-\bar{\theta}+\frac{eB\theta \bar{\theta}}{4\hbar}\right).
\end{eqnarray}
Using this relation, we obtain the commutation relation between the components of the NC cyclotron motion coordinates \cite{yan, Goerbig} given by 
$\hat{\eta}_{x}=\frac{1}{m\omega_{c}}\hat{\pi}_{y}$ and $\hat{\eta}_{y}=\frac{-1}{m\omega_{c}}\hat{\pi}_{x}$ 
\begin{eqnarray}
\label{e17}
\left[\hat{\eta}_{x},\hat{\eta}_{y}\right]=-i{l}_{B}^2
\end{eqnarray} 
where
\begin{eqnarray}
\label{e16}
l_{B}=\sqrt{\frac{\hbar}{eB}\left(1-\frac{eB\theta}{4\hbar}-\frac{\bar{\theta}}{eB\hbar}+\frac{\theta\bar{\theta}}{4\hbar^2}\right)}
\end{eqnarray}
is the magnetic length with corrections due to phase-space noncommutativity.

\noindent The commutation relation (\ref{e17}) gives us a nice way of accounting for the degeneracy of the Landau level. It implies that there is a Heisenberg uncertainty associated with the NC cyclotron variables $\hat{\eta}_{x}$, $\hat{\eta}_{y}$ and they obey 
\begin{eqnarray}
\label{e18}
\bigtriangleup\hat{\eta}_{x}\bigtriangleup\hat{\eta}_{y}=l_{B}^2.
\end{eqnarray}
Hence we can count the number of degenerate quantum states $N_{B}$  in a sample by dividing the sample area $A$ in the plane perpendicular to the magnetic field $\vec{B}$
by cell-area $l_{B}^2$ belonging to an individual state. This yields
\begin{eqnarray}
\label{e18a}
N_{B}&=&\frac{A}{\bigtriangleup\hat{\eta}_{x}\bigtriangleup\hat{\eta}_{y}}\nonumber\\
&=&\frac{eB}{\hbar}\left(1-\frac{eB\theta}{4\hbar}-\frac{\bar{\theta}}{eB\hbar}+\frac{\theta\bar{\theta}}{4\hbar^2}\right)^{-1}A\nonumber\\
&=&\frac{eB}{\hbar}\left(1+\frac{eB\theta}{4\hbar}+\frac{\bar{\theta}}{eB\hbar}+\frac{\theta\bar{\theta}}{4\hbar^2}+\frac{e^{2}B^2\theta^2}{16\hbar^2}+\frac{\bar{\theta^2}}{e^{2}B^2\hbar^2}\right)A+\mathcal{O}(\theta^{3}, \bar{\theta}^{3}, \theta^{2}\bar{\theta}, \theta\bar{\theta}^{2})\nonumber\\
\end{eqnarray}
The above expression reavels the effect of phase-space noncommutativity on the  degeneracy of Landau levels. 
To obtain the spectrum of the Landau problem in NC phase-space we write the Hamiltonian of the system in terms of the NC mechanical momenta as
\begin{eqnarray}
\label{e20}
\hat{H}=\frac{1}{2m}(\hat{\pi}_{x}^2+\hat{\pi}_{y}^2).
\end{eqnarray}
Introducing the ladder operators $\hat{a}$ and $\hat{a}^{\dagger}$ 
\begin{eqnarray}
\label{e21}
\hat{a}=\frac{\hat{\pi}_{x}-i\hat{\pi}_{y}}{\sqrt{2(eB\hbar-\frac{e^2B^2}{4}\theta-\bar{\theta}+\frac{eB\theta \bar{\theta}}{4\hbar})}} ~;\quad      \hat{a}^{\dagger}=\frac{\hat{\pi}_{x}+i\hat{\pi}_{y}}{\sqrt{2(eB\hbar-\frac{e^2B^2}{4}\theta-\bar{\theta}+\frac{eB\theta \bar{\theta}}{4\hbar})}}     
\end{eqnarray}  
where $\hat{a}$ and $\hat{a}^{\dagger}$ satisfy the commutation relation $[\hat{a}, \hat{a}^{\dagger}]=1$, the above Hamiltonian  can be expressed in the diagonal form as
\begin{eqnarray}
\label{e22}
\hat{H}=\hbar \tilde{\omega}_{c}\left(\hat{a}^{\dagger}\hat{a}+\frac{1}{2}\right)
\end{eqnarray}
where the original cyclotron frequency of the electron $\omega_{c}$ is replaced by
\begin{eqnarray}
\label{e22a}
\tilde{\omega}_{c}=\omega_{c}\left(1-\frac{eB}{4\hbar}\theta-\frac{1}{eB\hbar}\bar{\theta}+\frac{\theta \bar{\theta}}{4\hbar^2}\right)
\end{eqnarray}
The energy levels of the Landau problem in NC phase-space in terms of the familiar cyclotron frequency are therefore given by 
\begin{eqnarray}
\label{e25}
E_{n}=\hbar \omega_{c}\left(n+\frac{1}{2}\right)\left(1-\frac{eB}{4\hbar}\theta-\frac{1}{eB\hbar}\bar{\theta}+\frac{\theta \bar{\theta}}{4\hbar^2}\right); \quad n=0,1,2,...
\end{eqnarray} 
The above expession is exact to all orders of the spatial and momentum NC parameters. It clearly reveals that the Landau levels are altered by phase-space noncommutativity. In the $\bar{\theta}=0$ limit, the result agrees with that obtained by the path integral approach \cite{sg42}. Interestingly, the same reverse nature of the two leading order correction terms coming from the spatial and momentum sector, with respect to the change in the Background magnetic field, that was observed in the correction factor of the AB phase (\ref{e13a}) is present in the LL spectrum, and also in the level degeneracy.

To check the consistency of this result, we now want to compute the spectrum using commutative-equivalent Hamiltonian (\ref{e7}) written in terms of the commutative variables obtained by change of variables (\ref{e86}, \ref{e8zz}). Introducing the ladder operators involving the commutative phase-space variables (operators) $x$, $y$, $p_{x}$, $p_{y}$
\begin{eqnarray}
\label{e30a}
a_{x}=\frac{iap_{x}+bx}{\sqrt{2ab\hbar}}~; \quad a_{x}^{\dagger}=\frac{-iap_{x}+bx}{\sqrt{2ab\hbar}}
\end{eqnarray}
\begin{eqnarray}
\label{e31a}
a_{y}=\frac{iap_{y}+by}{\sqrt{2ab\hbar}}~; \quad  a_{y}^{\dagger}=\frac{-iap_{y}+by}{\sqrt{2ab\hbar}}
\end{eqnarray}
(where $a$ and $b$ are defined in eq.(\ref{e7})) which satisfy the commutation relations 
\begin{eqnarray}
\label{e30}
[a_{x},a_{x}^{\dagger}]=1=[a_{y},a_{y}^{\dagger}]
\end{eqnarray}
the transformed Hamiltonian (\ref{e7}) can be rewritten as
\begin{eqnarray}
\label{e34}
H=\frac{ab\hbar}{m}[(a_{x}^{\dagger} a_{x}+a_{y}^{\dagger} a_{y}+1)+i(a_{x}a_{y}^{\dagger}-a_{x}^{\dagger} a_{y})].
\end{eqnarray}
We further define the pair of operators
\begin{eqnarray}
\label{e32}
a_{+}=\frac{a_{x}+ia_{y}}{\sqrt{2}}~;\quad a_{-}=\frac{a_{x}-ia_{y}}{\sqrt{2}}
\end{eqnarray}
which satisfy the following commutation relations
\begin{eqnarray}
\label{e32a} 
[a_{+},a_{+}^{\dagger}]=1=[a_{-},a_{-}^{\dagger}]
\end{eqnarray}
to express the above Hamiltonian in the following diagonal form
\begin{eqnarray}
\label{e33}
\hat{H}=\frac{2ab\hbar}{m}\left(a_{-}^{\dagger}a_{-}+\frac{1}{2}\right).
\end{eqnarray}
Therefore the energy levels of the Landau problem can be immediately read off as
\begin{eqnarray}
\label{e33}
E_{n}&=&\frac{2ab\hbar}{m}\left(n+\frac{1}{2}\right)\nonumber\\
&=&\hbar \omega_{c}\left(n+\frac{1}{2}\right)\left(1-\frac{eB}{4\hbar}\theta-\frac{1}{eB\hbar}\bar{\theta}+\frac{\theta \bar{\theta}}{4\hbar^2}\right); \quad n=0,1,2,...
\end{eqnarray}
This expression for the NC corrected Landau levels is independent of the scaling parameter $\alpha$ appearing in the expression of generalized transformation (\ref{e86}) and  agrees with what we have obtained earlier using noncommutative variables (\ref{e25}), thus establishing the consistency of the two approaches.
\section{Conclusions}
In this paper we have investigated the effect of phase-space noncommutativity on the dynamics of a charged particle moving in a two dimensional plane in the presence of a background magnetic field perpendicular to the plane. 
This we do by mapping the problem in NC phase-space to an equivalent problem in the commutative plane by a generalized set of transformations (which we derive in this paper). In particular, we computed the phase-shift in Aharonov-Bohm effect and observed that the result gets modified by both the noncommutative parameters $\theta$ and $\bar{\theta}$ and also depends on the scaling parameter
appearing in the generalized transformations. The expression, with proper choice of the scaling parameter and terms retained upto first order in $\theta$ and $\bar{\theta}$, is found to agree with those in the existing literature obtained by canonical approach.
In the limit of vanishing momentum noncommutativity, the result to leading order in $\theta$ also agrees with that obtained by the path integral approach. The magnetic length, degeneracy of the Landau levels as well as the energy spectrum of the Landau problem is then computed using the NC phase-space variables and are found to get altered by phase-space noncommutativity. As a consistency check, the energy spectrum is also computed using an equivalent commutative version of the original theory and is found to match with the former one. This establishes the consistency of our framework and also the exactness of the transformation that has been used in the paper to relate the NC phase-space model with its commutative phase-space version.
 
\section*{Acknowledgement}
AS acknowledges the support by DST SERB under Grant No. SR/FTP/PS-208/2012.

\end{document}